\documentstyle[12pt]{article}
\begin{document}
\begin{titlepage}
\pagestyle{empty}
\baselineskip=21pt
\begin{center}
{\large{\bf
Ground-state Correlation Effects in Extended RPA Calculations
}} \end{center}
\vskip .1in
\begin{center}

A. Mariano and F. Krmpoti\'{c}

{\small\it Departamento de F\'\i sica, Facultad de Ciencias Exactas,}\\
{\small\it Universidad Nacional de La Plata, C. C. 67, 1900 La Plata,
Argentina}

{\small\it and}

A.F.R. de Toledo Piza

{\small\it Instituto de F\'\i sica, Universidade de S\~{a}o Paulo,\\
C.P. 20516, 01498 S\~{a}o Paulo, Brasil}

\end{center}
\vskip 0.5in
\centerline{ {\bf Abstract} }
We study normalization problems associated with use of  perturbatively
correlated ground-states in extended RPA schemes in the context of a specific
but typical example. The sensitivity of the results to the amount of $2p2h$
admixtures to the correlated ground state is also investigated in terms of a
modification of the standard perturbative approach.
\end{titlepage}
\baselineskip=18pt

\newpage
Some time ago Van Neck et al.\cite{Neck} pointed out that some annoying
numerical inconsistencies result from the evaluation of consistently derived
perturbative expressions in the context of the nuclear many-body problem.
Specifically, they pointed out that an often used procedure of evaluating
the number of nucleons perturbatively excited above the Fermi  level
(including normalization factors expanded to  the appropriate perturbative
order) leads to grossly overestimated results.  This happens due to the fact
that, as the perturbation adds a very large number of relatively small excited
$2p2h$ components to the $0p0h$ wave function, the relative weight of the
former in the perturbed wavefunction is typically large enough numerically
so that the  perturbatively expanded normalization becomes inadequate.
This type of difficulty affects  also linear response calculations done in
the context of the so called   extended second random phase approximation
(ESRPA) \cite{Arima}, which uses a perturbatively generated ground state
wavefunction with $2p2h$ admixtures, in addition to including two-body
operators in the structure of the phonons. In this note we work out an
example that shows that this numerical normalization  error tends in fact to
inflate significantly ESRPA strength distributions, as was  also pointed out
in Ref. \cite{Neck}. Moreover, considerable excess strength still remains over
the results obtained by using just the unperturbed ground state (second
random phase approximation (SRPA)) when one attends to the numerical
normalization problem. This excess strength appears to be related to the
relative importance of the $2p2h$ admixtures to the unperturbed ground state.

In order to explore this point we give also results obtained for a modified
ESRPA in which $2p2h$ ground state correlations are introduced by means of
the Brillouin-Wigner (BW) perturbation theory, which has the effect of reducing
appreciably their importance. This happens through the lowering of the
ground state energy produced by solving the  appropriate dispersion equation.
The resulting strength distribution, for reduced but still non negligible
$2p2h$ admixtures to the unperturbed ground state, comes out close to the
simple SRPA result.
We take these facts as an indication of enough
sensitivity of the calculated strengths to the correlation structure of the
ground-state so as to warrant the development and implementation in realistic
situations of better controlled extensions of the standard quasi-boson
 random-phase approximation.

We base our argument on the linear response $R(E)$ to an external field
$\hat{F}$, which admits the spectral representation
\begin{eqnarray}
R(E)=\sum_{\nu}\left[
 \frac{\langle 0 |\hat{F}| \nu \rangle \langle \nu |\hat{F}^{\dagger} | 0
\rangle}
 {E-E_{\nu}+i\eta}
-\frac{\langle 0 |\hat{F}^{\dagger}| \nu \rangle\langle \nu |\hat{F} | 0
\rangle}
{E+E_{\nu}-i\eta}
\right].
\label{1}
\end{eqnarray}
Here  $| 0 \rangle$ and $| \nu \rangle$ are the exact ground state and excited
eigenstates of the full hamiltonian $\hat{H}$.
The excited states $| \nu \rangle$
can be written as
\begin{eqnarray}
| \nu \rangle= \Omega_{\nu}^{\dagger}| 0 \rangle ;
\hspace{2mm}\Omega_{\nu}^{\dagger} = \sum_{i} X_{i}^{\nu} C_{i}^{\dagger}
- \sum_{j} Y_{j} ^{\nu}C_{j},
\label{2}
\end{eqnarray}
where the set $\{C_i,C_i^{\dagger}\}$  constitutes a complete operator basis.
In the SRPA this set is restricted to one and two particle-hole annihilation
and creation operators out of the Hartee-Fock (HF) ground state $|HF \rangle$
and the coefficients $X_i^{\nu}$ and $Y_i^{\nu}$ are determined from the
equations of motion \cite{Rowe}
\begin{eqnarray}
\langle HF
|\left[\Omega_{\mu},\left[\hat{H},\Omega_{\nu}^{\dagger}\right]\right]|HF
\rangle
= E_{\nu}\langle HF |\left[\Omega_{\mu},\Omega_{\nu}^{\dagger}\right]|HF
\rangle
= E_{\nu} \delta_{\mu \nu}.\label{3}
\end{eqnarray}
The ESRPA hinges on the idea that the inclusion of $2p2h$ operators among
the $C_i$ requires a modification of the quasi-boson
approximation in which the HF ground state is allowed to include perturbative
$2p2h$ correlations. Hence, in evaluating Eq. (\ref{3}) one uses in this case
a ground state of the form
\begin{eqnarray}
|\tilde{0}\rangle = c_0\left[ |HF \rangle + \sum_{2_0} c_{2_0}
|2_0 \rangle\right],\label{4}
\end{eqnarray}
where
the amplitudes $c_{2_0}$ are evaluated in the first order
Rayleigh-Schr\"{o}dinger (RS) perturbation theory, i.e.,
\begin{eqnarray}
c_{2_{0}}=\frac{\langle 2_0 |\hat V |HF \rangle}
{-E_{2_{0}}}.\label{5}
\end{eqnarray}
Here $2_0\equiv(p_1p_2h_1h_2)_0$ indicates $2p2h$ excitations with
independent-particle energy $E_{2_0}$, $\hat{V}$ is the residual interaction,
and $c_0$ is a normalization factor.
Since Eq. (\ref{4}) is thought of as a perturbatively generated expression,
$c_0$ is generally set equal  to 1.

In general Eq. (\ref{3}) leads to a secular problem of the form
\begin{eqnarray}
{\cal A} {\cal X}^\nu = E_{\nu} {\cal N} {\cal X}^{\nu},
\label{6}
\end{eqnarray}
with
\begin{eqnarray}
{\cal A} = \left(\begin{array}{cc} A & B \\ B^{\ast} &
A^{\ast}\end{array}\right),
\hspace{5mm}
{\cal X}^{\nu} = \left(\begin{array}{l} X^{\nu}\\Y^{\nu}\end{array}\right),
\hspace{5mm}
{\cal N} = \left(\begin{array}{cc} N & 0\\ 0 &-N^{\ast}\end{array}\right).
\label{7}
\end{eqnarray}
where the submatrices $A$, $B$ and $N$ given by
\begin{eqnarray}
A_{i,j} = \langle
\tilde{0}|\left[C_i,\left[\hat{H},C_{j}^{\dagger}\right]\right]
| \tilde{0}\rangle,
\hspace{2mm}
B_{i,j} = \langle \tilde{0}|\left[C_i,\left[\hat{H},C_{j}\right]\right]
| \tilde{0}\rangle,
\hspace{2mm}
N_{i,j} = \langle \tilde{0}|\left[C_i,C_j^{\dagger}\right]| \tilde{0}\rangle.
\label{8}
\end{eqnarray}
Furthermore, one can write Eq. (\ref{1}) in a representation independent
form as
\begin{eqnarray}
R(E) = {\cal F}^{\dagger} ( E {\cal N} + i\eta {\cal I}-{\cal A})^{-1} {\cal
F},
\label{9}
\end{eqnarray}
where the matrix ${\cal F}$ represents  the operator $\hat {F}$ and is defined
as
\begin{eqnarray}
{\cal F}\equiv \left(\begin{array} {ll} F^{A} \\ F^{B} \end{array}\right),
\;\;\mbox{ with}
\left\{ \begin{array}{l}
F^A_i=\langle \tilde{0}|\left[ C_i,\hat{F}\right]| \tilde{0}\rangle,\\
\vspace{2mm}
F^B_i = F_i^{A\ast}(\hat{F} \rightarrow \hat{F}^{\dagger}).
\end{array}\right.
\label{10}
\end{eqnarray}
Eq. (\ref{9}) can be reduced with the help of projection operators $P$ and
$Q$ onto subspaces involving $1p1h$ and $2p2h$ excitations respectively.
One gets
\begin{equation}
R(E) = \tilde{\cal{F}}_P^{\dagger}(E){\cal G}_P(E) \tilde{\cal F}_P(E)
+ {\cal F}_Q^{\dagger}{\cal G}_Q(E){\cal F}_Q,
\label{11}
\end{equation}
where
\begin{equation}
{\cal G}_P(E)=\left[E {\cal N}_P+i\eta {\cal I}_P-{\cal A}_P
-\left({\cal A}_{PQ}-{\cal N}_{PQ}E \right)
{\cal G}_Q(E) \left({\cal A}_{QP}-{\cal N}_{QP}E \right)\right]^{-1},
\label{12}
\end{equation}
with
\begin{equation}
{\cal G}_Q(E)=\left[E{\cal N}_Q+i\eta {\cal I}_Q-{\cal A}_Q \right]^{-1},
\label{13}
\end{equation}
and
\begin{equation}
\tilde{\cal F}_{P}(E)={\cal F}_P-{\cal N}_{PQ}{\cal F}_Q
+{\cal A}_{PQ}{\cal G}_{Q}(E){\cal F}_Q.
\label{14}
\end{equation}
When using the ESRPA some more complicated two-body effects are trimmed
by keeping terms up to second order in $\hat{V}$ for the forward sector
within the P space, terms linear in $\hat{V}$ for the backward sector within
the P space and for the coupling between the P and Q spaces, and only
terms of zeroth order within the Q space.
The usual argument (see e.g. Ref. \cite{Arima}) for this procedure involves
again the
limitations stemming from the perturbative dressing of the ground state,
Eq. (\ref{4}).

Finally, we set up a modified extended second RPA (MESRPA) in which
Eq. (\ref{4}) is replaced by the corresponding expression obtained from
the BW perturbation theory. This in fact coincides with the
ESRPA result but the coefficients $c_{2_0}$ are now given as
\begin{eqnarray}
c_{2_{0}}= \frac{ \langle 2_0| \hat V |HF \rangle}{E_{0}- E_{2_0} },
\label{15}
\end{eqnarray}
where the ground state energy $E_0$ is the lowest
solution of the secular equation
\begin{eqnarray}
E_0 = \sum_{2_0} \frac{\vert \langle 2_0| \hat V |HF
\rangle\vert^2}{E_0-E_{2_0}}.
\label{16}
\end{eqnarray}
This leads to increased energy denominators in Eq. (\ref{15}) and hence to
reduced $2p2h$ admixtures to the HF ground state. One obtains in this way
\begin{equation}
N_{ij}=\delta_{ij} + \Delta N_{ij}
\label{17}
\end{equation}
where $i\equiv ipih$ and the nonzero $\Delta N_{ij}$ are just
\begin{eqnarray}
& &\Delta N_{11'}= \vert c_0 \vert ^2
\sum_{2_0,2_0'}c_{2_0}^{\ast}c_{2_{0}'} \langle 2_0|\hat{D}_{11'}
|2_{0}'\rangle,
\label{18}
\end{eqnarray}
where
$\hat{D}_{11'}=\left[\hat{C}_1,\hat{C}_{1'}^{\dagger}\right]-\delta_{11'}$.
(Note that within the quasi-boson approximation $\hat{D}_{11'}\equiv 0$).
The explicit result for the matrix element
$\langle 2_0|\hat{D}_{11'} |2_0'\rangle$ is
\begin{eqnarray}
&&\langle (p_1ph_1h_2)_0|\hat{D}_{ph,p'h'}|(p'_1p'_2h'_1h'_2)_0\rangle
=-\left[1+P(h_1,h_2)P(h_{1'},h_{2'})\right]\\
\nonumber
&\times& \left[\delta_{p,p'}\delta_{h_1,h'} P^-(h,h_2)P^-(p_1,p_2)
\delta_{h_{1'},h}\delta_{h_2,h_{2'}}\delta_{p_2,p_{2'}}\delta_{p_1,p_{1'}}
\right] + p \leftrightarrow h ,
\label{19}
\end{eqnarray}
where $P^-(i,j)\equiv [ 1 - P(i,j)]$, while the operator $P(i,j)$
exchanges the arguments $i$ and $j$.
The matrix elements of ${\cal A}$ are
\begin{equation}
A_{ij}=\delta_{ij}E_j + V_{ij} +\Delta A_{ij},
\label{20}
\end{equation}
where $V_{ij}\equiv \langle i\hat{V}|j\rangle$ and
the nonzero matrix elements $\Delta A_{ij}$ are:
\begin{equation}
\Delta A_{11'}=\vert c_0 \vert ^2
\sum_{2_0,2_{0}'}(E_1-E_{2_0}+E_0)c_{2_0}^{\ast}c_{2_{0}'}
\langle 2_0|\hat{D}_{11'}|2_{0}'\rangle .\label{21}
\end{equation}
Finally the matrix elements of ${\cal F}$ are:
\begin{equation}
F_i^A =  \left\{ \begin{array}{l}
f_1 + \sum_{1'} \Delta N_{11'}f_{1'} \hspace{5mm} \mbox{for $i=1$}
\\
 c_0   \sum_{2_0} c_{2_0} f_{22_0}\hspace{11mm}\mbox{for $i=2$,}
\end{array}\right.
\label{22}
\end{equation}
where
\begin{equation}
f_1 \equiv \langle 1|\hat{F}|HF \rangle, \hspace{2mm} \mbox{and} \hspace{2mm}
f_{22_0}\equiv \langle 2|\hat{F}|2_0 \rangle.
\label{23}
\end{equation}
Note that the corresponding ESRPA quantities are obtained by setting $c_0=1$
and $E_0=0$ in the MESRPA expressions.

We next give  numerical results  for the GT resonance
($\hat{F}_{\pm}\equiv \vec{\sigma} t_{\pm}$) in $^{48}Ca$ using the MY3 force
\cite{Ber} in the $0\hbar\omega-3\hbar\omega$ oscillator space.
Four different ways of handling the nuclear ground state will be compared.
The first one is just the plain SRPA in which the equations of motion (\ref{3})
are evaluated with the HF ground state. We give also results for the ESRPA
(for which the normalization coefficient $c_0$ is set equal to 1), for a
normalized version of this approximation (NESRPA) in which $c_0$ is determined
so that $\langle {\tilde{0}}|{\tilde{0}}\rangle =1$ with the $c_{2_0}$
coefficients given by Eq.
(\ref{5}), and finally for the MESRPA, which uses a normalized ground state
with amplitudes $c_{2_0}$ evaluated using Eq. (\ref{15}). In order to obtain
smooth strength functions $S(E) \equiv -\frac{1}{\pi}Im~R(E)$ with $R(E)$
given by Eq. (\ref{11}), the energy variable is taken to be complex:
$E \rightarrow E+i\Delta$, with $\Delta=1$ MeV for the $1p1h$ and
$\Delta=3$ MeV for the $2p2h$ subspace respectively. Solving the dispersion
equation Eq. (\ref{16}) gives $E_0=-29$ Mev, which amounts to about 8\% of the
experimental ground state binding energy.
The results are shown in Fig. 1 and in Table 1 below.
The positive branch of  the GT sum rule
${\cal S}_+ - {\cal S}_- =3(N-Z)$ with
${\cal S}_{\pm}\equiv \int S_{\pm}(E)dE$ is divided into a low energy part
${\cal S}_+^<$ ($ E < 20$ Mev) and a high energy part
${\cal S}_+^>$ ($ E > 20$ Mev). These quantities are the relevant ones for the
problem of the quenching  of  GT strength. In the usual RPA the low energy
part ${\cal S}_{+}^{<}$ essentially exhausts the sum rule. When $1p1h-2p2h$
coupling is introduced via the SRPA, 30\% of this strength is shifted to the
high energy region. This amount is somewhat reduced when ground state
correlations are introduced via the ESRPA.
This results from the combined effect of the Q-space part of Eq. (\ref{11})
and of the interference effects generated by the last term of Eq. (\ref{14})
\cite{Nis1}. Furthermore one gets now also  a contribution in the negative
branch  ${\cal S}_{-}$ so that ${\cal S}_+^<$ increases to 82\%
(third line in Table 1).
As shown in the first two columns of Table 1, the ground state wavefunction
involved in the derivation of the ESRPA expressions has a serious
normalization problem. In the NESRPA this is fixed by suitably reducing the
value of $c_0$. This has only a relatively small effect on ${\cal S}_+^<$ and
reduces both ${\cal S}_+^>$ and the negative branch contribution ${\cal S}_{-}$
(fourth line of Table 1).
When ground state correlations are introduced via the MESRPA, on the other
hand, the percentage of $2p2h$ admixtures to the ground state is reduced from
62\% to 31\% while the strength distribution becomes quite similar to the
simple SRPA result.
This last feature indicates important sensitivity to the amount of ground
state correlations which therefore, as stated above, deserves a more
controlled treatment.
It is worth noticing that,  when the interaction among the $2p2h$
configurations is neglected, the BW approximation  for the ground state
wavefunction coincides with the diagonalization procedure \cite{Boh}. This
means that the $0p0h-2p2h$ coupling in the initial nucleus is treated at the
same footing as the $1p1h-2p2h$ coupling in the final nucleus.
Thus we feel that, in the context of the present calculations, it is
more consistent to use the BW perturbation theory than the RS one.

Even though the present discussion has been limited to one specific case
involving the Gamow-Teller response within the extended RPA, the observed
trends should apply also to other schemes which include ground state
correlation effects perturbatively, both for this \cite{Ber} and for other
types of response functions, notably  the longitudinal and transverse inclusive
responses in quasi-free electron scattering \cite{Alb,Tak,Mar}. In all cases
ground-state  normalization is numerically important and sensitivity to the
amount of $2p2h$ correlations should be  high, so that a moderate
reduction of the $2p2h$ ground state component will lead to results which
are not far from those obtained in the simple SRPA.
\newpage

\begin{center} { \large ACKNOWLEDGMENTS} \end{center}
AM and FK are fellows of the CONICET from Argentina.
AFRTP is indebted to CCInt-USP and to the UNLP  for financial help.

\newpage
\begin{table}
\begin{center} { \large TABLES }
\caption{Ground state normalization factor, summed weights of $2p2h$
components (see Eq. (\protect \ref{4})), GT integrated strength ${\cal S}_+$ in
the
resonance region $(<)$, above it $(>)$, and total GT strengths ${\cal S}_+$
and ${\cal S}_-$.
The first column identifies the approximation scheme. Strengths are given in
percent of $3(N-Z)$.}
\label{table1}

\bigskip

\begin{tabular}{lcccccc}\hline
&$|c_0|^2$ & $\sum_{2_0} |c_{2_0}|^2$ & ${\cal S}_{+}^{<}$ &
${\cal S}_{+}^{>}$ & ${\cal S}_{+}$ & ${\cal S}_{-}$\\
\hline
$RPA$    & 1     & 0   & 100  & 0   & 100 & 0 \\
$SRPA$   & 1     & 0   &  70  & 30  & 100 & 0 \\
$ESRPA$  & 1     & 1.60  & 82 & 27  & 109 &  9 \\
$NESRPA$ & 0.38 & 1.60  & 80  & 23.5 & 103.5& 3.5  \\
$MESRPA$ & 0.69  & 0.45 & 72.8 & 28  & 100.8& 0.8 \\
\hline
\end{tabular}
\end{center}
\end{table}

\bigskip

\begin{figure}
\begin{center} { \large FIGURES}\end{center}
\caption{Folded Gamow-Teller strength distributions in \protect $^{48}Ca$
for different approximation schemes.}
\label{fig}
\end{figure}

\begin{thebibliography}{99}
\bibitem {Neck} D. Van Neck, M. Waroquier, V. Van der Sluys and
J. Ryckebusch, Phys.\ Lett.\ {\bf 274B}, 143 (1992).
\bibitem {Arima} K. Takayanagi, K. Shimizu and A. Arima, Nuc.\ Phys.\
{\bf A477}, 205 (1988).
\bibitem {Nis1} S. Dro\.{z}d\.{z}, S. Nishizaki, J. Speth and J. Wambach,
Phys.\ Rep. \ {\bf 197}, 1 (1990).
\bibitem {Rowe} D.J. Rowe, Nuclear Collective Motion (Methuen, London 1970).
\bibitem{Dro1}S. Dro\.{z}d\.{z}, V. Klemt, J. Speth and J. Wambach, Phys. Lett.
 {\bf 166B}, 253 (1986).
\bibitem {Boh} A. Bohr and B. Mottelson, {\it Nuclear Structure} \ Vol. I,
p. 302 (Benjamin, New York).
\bibitem {Ber} G.F. Bertsch and I. Hamamoto, Phys.\ Rev.\
{\bf C26}, 1323 (1982).
\bibitem{Alb} W.M. Alberico, M. Ericson and A. Molinari, Ann. Phys. (NY)
{\bf 154}, 356 (1984).
\bibitem {Tak} K. Takayanagi, Phys. \ Lett. \  {\bf 233B}, 271 (1989).
\bibitem {Mar} A. Mariano, E. Bauer, F. Krmpoti\'{c}, and
A.F.R de Toledo Piza, Phys. \ Lett. \ {\bf 268B}, 332 (1991).
\end{thebibliography}
\end{document}